\documentclass[12pt]{iopart}
\input epsf
\newcommand{\vk}{{\bf{k}}}
\newcommand{\vr}{{\bf{r}}}
\newcommand{\omk}{\omega_{\vk}}
\newcommand{\be}{\begin{equation}}
\newcommand{\ee}{\end{equation}}
\newcommand{\bea}{\begin{eqnarray}}
\newcommand{\eea}{\end{eqnarray}}
\newcommand{\non}{\nonumber\\}
\newcommand{\no}{\nonumber}
\newcommand{\lbr}{\left\{ }
\newcommand{\rbr}{\right\} }
\newcommand{\lp}{\left ( }
\newcommand{\rp}{\right ) }
\newcommand{\lb}{\left [ }
\newcommand{\rb}{\right] }
\newcommand{\ld}{\left. }
\newcommand{\rd}{\right. }
\newcommand{\cM}{{\cal M}}
\newcommand{\rhok}{\rho_{\vk}}
\newcommand{\rhomk}{\rho_{-\vk}}
\newcommand{\sumk}{\sum_{\vk}}
\newcommand{\gs}{\gamma\delta}
\newcommand{\bcM}{\bar{\cM}}

\usepackage{iopams,epsf}
\begin{document}

\title[]{Ab initio study
of the vapour-liquid critical point of a symmetrical binary fluid mixture  }

\author{O V Patsahan, M P Kozlovskii and R S Melnyk
\footnote[3]{Author to whom any correspondence 
should be addressed. E-mail address: romanr@icmp.lviv.ua}}

\address{\ Institute for Condensed Matter Physics,
1 Svientsitskii Street, Lviv 290011, Ukraine }

\begin{abstract}
A microscopic approach to the investigation of the behaviour 
of a symmetrical binary fluid mixture in the vicinity 
of the vapour-liquid critical point is proposed. It is shown 
that the problem can be reduced to the calculation of the partition 
function of a 3D Ising model in an external field. For a square-well 
symmetrical binary mixture we calculate the parameters of the 
critical point as functions of the microscopic parameter $r$ measuring the 
relative strength of interactions between the particles of dissimilar and 
similar species. The calculations are performed at intermediate 
($\lambda=1.5$) and moderately long ($\lambda=2.0$) intermolecular 
potential ranges. The obtained results agree well with the ones of computer 
simulations.
\end{abstract}

\pacs{05.70.Jk}


\maketitle

\section{Introduction}
\noindent
Binary mixtures in contrast to their constituent components can exhibit 
three different types of two-phase equilibria: vapour-liquid, liquid-liquid 
and gas-gas \cite{1,2}. The possibility of the realization 
of these phenomena and their priority depend both on the external conditions 
and microscopic parameters of a mixture. The study of tne influence of
interparticle interactions on the critical properties of a binary mixture 
is an interesting and actual problem. During the last decade this 
problem has been intensively studied by integral equation 
methods \cite{3}-\cite{9}.
However, this approach, although reproduces different phase diagram 
types by varying the microscopic parameters, gives only a qualitative 
picture of the phenomena under consideration. On the other hand, the 
critical properties of simple fluids and binary mixtures have been 
studied recently using Monte Carlo (MC) simulations \cite{10}-\cite{14}.
In \cite{14} the vapour-liquid critical 
temperature was calculated for the symmetrical mixture of hard spheres 
interacting via the square-well potentials. Thus, it is interesting to test 
a theory using such a simple binary fluid model.

In the present paper we propose a microscopic approach to the study of 
the vapour-liquid critical point of a symmetrical binary mixture. This 
approach is based on the method of collective variables (CV). This method 
was developed in \cite{15} and appears to be successful in describing 
the second order phase transition of the 3D Ising model \cite{16} and 
the vapour-liquid critical point of a one-component fluid \cite{17}.  
On the basis of this approach both universal and non-universal quantities 
were obtained.

In \cite{18} the CV method with a reference system (RS) was 
generalized for the case of a grand canonical ensemble for a multicomponent 
continuous system. Using this approach the phase diagram of the symmetrical 
mixture was examined within the framework of the Gaussian 
approximation \cite{19}-\cite{21}.

In this paper we determine an explicit form of the effective 
Ginsburg-Landau-Wilson (GLW) Hamiltonian of the symmetrical binary mixture 
in the vicinity of the vapour-liquid critical point. Then we integrate 
the functional of the grand partition function by the use of the 
layer-by-layer integration method proposed in \cite{16} for the 3D Ising 
model. As a result of this integration one obtains recursion relations for 
the coefficients of the GLW Hamiltonian. The analysis of these relations 
yields an equation for $T_c$. Here we avoid extensive consideration of the
results pertaining to the Ising model and call the readers' attention to 
\cite{16} where this problem was studied in detail. The method which 
we describe here yields the same critical exponents as in \cite{22} 
(see table 1).

\begin{table}[htbp]
\caption{Values of the critical exponents and the ratios of the 
critical amplitudes for the 3D Ising model obtained within the framework of 
the CV method.}
\begin{indented}
\item[]
\begin{tabular}{@{}llllll}
\br
$~\nu~$ &$~\alpha~$ &$~\beta~$ &$~\gamma~$ 
&$~A^+/A^-~$&$~\Gamma^+/\Gamma^-~$\\
\mr
0.637 &0.088 & 0.319 & 1.275 & 0.435 & 6.967\\
\br
\end{tabular}
\end{indented}
\end{table}

The paper is organized as follows. We give a functional representation of the
grand partition function of a two-component continuous system in section 2
and appendix A. In section 3 we construct the basic density measure 
(GLW Hamiltonian) with respect to the CV which include a 
variable corresponding to the order parameter. In this section we also 
present the basic ideas of the method of the partition function 
integration in the vicinity of the critical point. In section 4 we apply 
our formalism to calculating the critical characteristics (temperature and 
density) of the binary square-well symmetrical mixture. The obtained results 
are discussed and compared with the MC simulation data reported recently by 
N.B.Wilding \cite{14} and E.de Miquel \cite{13}.

\section{Functional representation  of the grand partition function of a 
binary mixture }
\indent
Let us consider a binary fluid mixture consisting of $N_a$ particles of 
species "a" and $N_b$ particles of species "b". The system is 
in volume $V$ at temperature $T$.
Let us assume that an interaction in the system has a pairwise additive 
character. The interaction potential between $\gamma$ particle at ${\vr}_i$
and $\delta$ particle at ${\vr}_j$ can be expressed as a sum of two terms: 
\bea
U_{\gs}(|{\vr}_i-{\vr}_j|)=\Psi_{\gs}
(|{\vr}_i-{\vr}_j|)+\Phi_{\gs}(|{\vr}_i-{\vr}_j|),
\label{2.0}
\eea
where $\Psi_{\gs}(r)$ is a potential of a short-range 
repulsion and  $\Phi_{\gs}(r)$ is an attractive part of the potential 
which dominates at large distances. 

A functional of the grand partition function (GPF) of the binary 
homogeneous system in the CV method with a RS can be represented as a 
product of two factors (see Appendix~A):

\be
~~~~~~~~~~~~~~~~~~~~~~~~~~~\Xi = \Xi_0 \Xi_1,
\label{2.1}
\ee
where $\Xi_0$ is the GPF of the RS which we suppose to be 
known. $\Xi_1$ is the part of the GPF which is written in the CV space:

\bea
\Xi_1 & =& \int (d\rho)(dc)
  \exp \lbr \beta \mu_1^+\rho_0+\beta\mu_1^-c_0-\rd
 \frac{\beta}{2} \sumk
  \lb \tilde V(k) \rhok\rhomk \rd + \non
& + & \ld\ld 2\tilde U(k)\rhok c_{\vk} +
  \tilde W(k)c_{\vk} c_{-\vk} \rb\rbr J(\rho,c).
\label{2.2}  
\eea
Chemical potentials $\mu_1^+=\frac{1}{\sqrt 2}(\mu_1^a+\mu_1^b)$ 
and $\mu_1^-=\frac{1}{\sqrt 2}(\mu_1^a-\mu_1^b)$ are 
determined from the conditions:

\numparts	
\bea
\frac{d ln \Xi_1}{d \beta\mu_1^+} & = & <N_a> + <N_b> =<N>\\
\frac{d ln \Xi_1}{d \beta\mu_1^-} & = & <N_a> - <N_b>.
\label{2.3}
\eea
\endnumparts
Functions $\tilde V(k), \tilde W(k)$ and $\tilde U(k)$ are 
combinations of Fourier transforms of the initial interaction potentials 
$\tilde{\Phi}_{\gs}(k)$:

\bea
\tilde V(k)& =& (\frac{\beta^{-1}}{2})\lb \alpha_{aa}(k)+\alpha_{bb}(k)+
  2\alpha_{ab}(k)\rb,\non
\tilde U(k)& =& (\frac{\beta^{-1}}{2})\lb
  \alpha_{aa}(k)-\alpha_{bb}(k)\rb,\\
\tilde W(k)& =& (\frac{\beta^{-1}}{2})\lb \alpha_{aa}(k)+\alpha_{bb}(k)-
  2\alpha_{ab}(k)\rb.\no
\label{2.4}  
\eea

\bea
J(\rho,c) & = & \int (d\nu)(d\omega) \exp \lbr i2\pi \sumk
  (\omk\rhok+\nu_{\vk}c_{\vk})+\rd\non
& + & \ld \sum_{n\geq 1}\sum_{i_n\geq 0} D_n^{(i_n)}(\omega,\nu)\rbr
\label{2.5}
\eea
is a transition Jacobian to the CV $\rhok, c_{\vk}$ averaged on 
the RS, variables $\omk, \nu_{\vk}$ are conjugated to variables 
$\rho_{\vk},c_{\vk}$, respectively.

\bea
D_n^{(i_n)}(\omega,\nu) & = & \lb \frac{(-i2\pi)^n}{n!}\rb
\lp\frac{1}{2}\rp^{n/2} \sum_{k_1...k_n} \cM_n^{(i_n)}(0,...,0)\times\non
& \times & \nu_{\vk_1}\ldots\nu_{\vk_{i_n}}
\omega_{\vk_{i_n+1}}\ldots\omega_{\vk_n}
\delta_{\vk_1+\ldots+\vk_n}.
\label{2.6}
\eea
Index $i_n$ indicates the number of variables $\nu_{\vk}$ in the 
cumulant expansion (\ref{2.6}). Cumulants 
$\cM_n^{(i_n)}(0,...,0)$ are linear combinations of the initial 
cumulants $\cM_{\gamma_1...\gamma_n}(0,...,0)$ $(\gamma_i=a,b)$ (see 
Appendix B).

We consider a symmetrical binary fluid mixture (SBFM), 
i.e. a system in which the two pure components "a" and "b" are identical 
and only interactions between the particles of dissimilar species differ.  
Notwithstanding its simplicity, the SBFM exhibits all the
three types of two-phase equilibria which are observed in real binary 
fluids, namely: vapour-liquid, liquid-liquid and gas-gas equilibria. 
For the SBFM $\tilde U(k)=0$ in (\ref{2.2}) and there are only terms with 
even indices $i_n$ in the cumulant expansion (\ref{2.6}) \cite{20}. 

\section{The method}
\noindent
As it was already shown \cite{20}, the phase diagram of the SBFM consists 
of three ranges (see figure 1): (1) gas-gas separation and  
vapour-liquid phase transitions;  (2) vapour-liquid and liquid-liquid phase 
transitions; (3) vapour-liquid phase transition only. 
\begin{figure}
\centerline{
\epsfbox{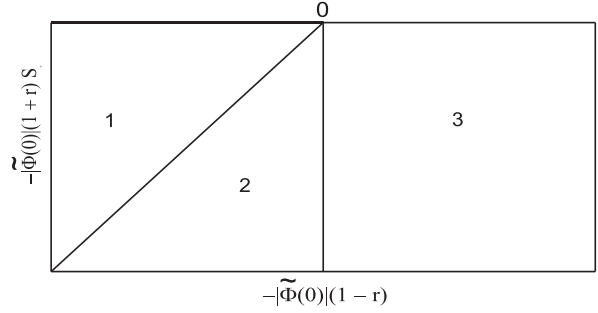}}
\caption{Three phase regions of the symmetrical mixture depending 
on the microscopic parameters: (1) gas-gas and vapour-liquid phase 
transitions ($T_c^{g-g} > T_c^{v-l}$); (2) 
vapour-liquid and liquid-liquid phase transitions ($T_c^{v-l} > T_c^{l-l}$);
 (3) vapour-liquid phase transition only. $S_+$ is the structure factor of 
 the reference system.}
\label{f1}
\end{figure}

The order of 
priority of the vapour-liquid and separation phase transitions depends on 
both the external conditions and the microscopic properties of the system. 
There exist two critical temperature branches in such a system: branch 
$(T_c^{v-l})$ connected with the  variable $\rho_0$ and branch 
$(T_c^{sep})$ connected with the variable $c_0$ \cite{20}. All the 
thermodynamic functions of the SBFM are symmetrical with respect to the 
concentration x=0.5 and have an extremum at this point \cite{1,19}. 
The concentration x=0.5 is a critical one for this model mixture.

 In this paper we consider a symmetrical fluid mixture within range 2 
or 3 of its phase diagram. 

 In the case of the SBFM the variables $\rho_0$ and $c_0$ are connected  
with the order parameters for the vapour-liquid and separation phase  
transitions, respectively \cite{20}. This fact allows us to separate 
CV $\rho_{\vk}$ and $c_{\vk}$ into  
essential and non-essential ones depending on the  
phase transition considered. Since we are interested in the vapour-liquid 
critical point, we can consider   
CV $c_{\vk}$ (and $\nu_\vk$) to be non-essential (CV 
$c_{\vk}$ do not contain a variable connected with the order parameter, 
the coefficients standing at the second power of $c_\vk$ (and $\nu_\vk$) 
are negative) and we can integrate over $c_\vk$ (and $\nu_\vk$) with 
the Gaussian density measure. In respect to CV $\rho_\vk$ it is 
necessary to construct the basic density measure taking into account 
higher powers of $\rho_\vk$ (we shall consider a $\rho^4$ model).

As a result of integrating in (\ref{2.2})-(\ref{2.6}) over variables 
$c_\vk$ (and $\nu_\vk$) we obtain for the GPF:

\be
\Xi  =  \Xi_0\Xi_G^c \int (d\rho) \exp \lbr \beta \mu_1^+\rho_0 -
\frac{\beta}{2}\sum_{\vk}\tilde V(k)\rhok\rhomk\rbr J(\rho),~
\label{3.1}
\ee
where
\bea
\Xi_G^c = \prod_{\vk} \frac{1}{\sqrt{1+\beta\tilde W(k)\cM_2^{(2)}(0)/2}},
\label{3.2}
\eea

\bea
J(\rho) & =& \int (d\omega) \exp \lbr i2\pi \sum_{\vk}\omk\rhok
  +  \sum_{n\geq 1}^4\frac{(-i2\pi)^n}{n!}
\lp\frac{1}{2}\rp^{n/2}\times\rd\non
&\times&\ld  \sum_{\vk_1...\vk_n} \cM_n(0)
  \omega_{\vk_1} \ldots \omega_{\vk_n}\delta_{\vk_1+...+\vk_n}\rbr,
\label{3.3}  
\eea
\bea
\cM_n(0)=\cM_n^{(0)}(0)+\Delta\cM_n.~
\label{3.4}
\eea
$\Delta \cM_n$ are the corrections obtained as the result of integration 
over variables $c_{\vk}$:
\bea
\Delta \cM_1 & = &
 \frac{\cM_3^{(2)}(0)}{12}\frac{1}{<N>}\sum_{\vk}\tilde g(|\vk|),\non
\Delta \cM_2 & = & \frac{\cM_4^{(2)}(0)}{12}\frac{1}{<N>}\sum_{\vk}\tilde
g(|\vk|)+\non
&+&\frac{(\cM_3^{(2)}(0))^2}{72}\frac{1}{<N>^2}\sum_{\vk}\tilde
g(|\vk|)\tilde g(|\vk_1-\vk|),\non
\Delta \cM_3 & = &
\frac{\cM_3^{(2)}(0)\cM_4^{(2)}(0)}{48}\frac{1}{<N>^2}\sum_{\vk}\tilde
g(|\vk|)\tilde g(|\vk_1-\vk|)+\non
&+&\frac{(\cM_3^{(2)}(0))^3}{6}\frac{1}{<N>^3}\sum_{\vk}\tilde g(|\vk|)
  \tilde g(|\vk_1+\vk|)\tilde g(|\vk_2-\vk|),\non
\Delta \cM_4 & = &  
\frac{(\cM_4^{(2)}(0))^2}{96}\frac{1}{<N>^2}\sum_{\vk}\tilde 
g(|\vk|)\tilde g(|\vk_1-\vk|)+\non
&+& (\frac{\cM_3^{(2)}(0)}{3!})^4\frac{1}{<N>^4}\sum_{\vk}\tilde
   g(|\vk|)\tilde g(|\vk_1+\vk|) \times \non
&\times&   \tilde g(|\vk_2-\vk|)\tilde g(|\vk_3+\vk_1+\vk|),
\label{3.5}
\eea
where
\bea
\tilde g(k) & = & -\frac{\beta <N> \tilde W(k)}
{\frac{1}{2}\beta \tilde W(k)\cM_2^{(2)}(0)+1}.
\label{3.6}
\eea

In figure 2 the  typical behaviour of the potenial $\tilde V(k)/|\tilde 
V(0)|$ is shown. 
\begin{figure}[htbp]
\centerline{
\epsfbox{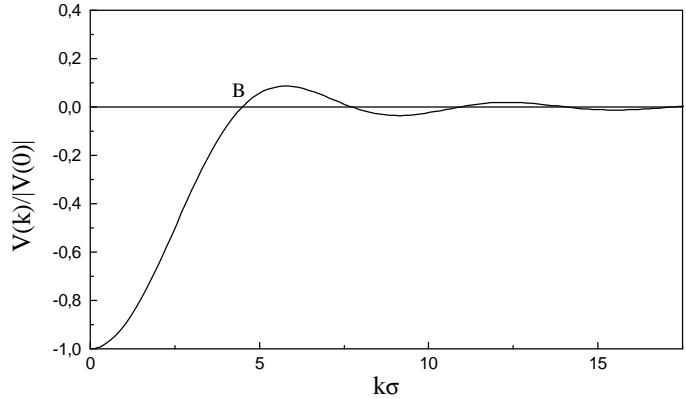}}
\caption{The behaviour of the Fourier transform $\tilde 
V(k)/|\tilde V(0)|$ of the attractive part of the interaction potential 
V(r).}
\label{f2}
\end{figure}

Let us further assume that 
$\tilde V(k)=0$ at $|\vk|>B$. Then, integration in (\ref{3.1}) over 
$\rhok$ with $|\vk|>B$ leads to $\delta$ - functions and the expression 
for $\Xi$ contains only the sums over $\vk$ with  $|\vk|\leq B$. 

We consider a set of $\vk$ vectors, $|\vk|\leq B$, as  
corresponding to the sites of a reciprocal lattice conjugated to a certain 
block lattice $\{r_l\}$ with $N_B$ block sites in the periodicity volume $V$:
\be
<N_B> = \frac{V}{C^3} = \frac{V}{(\pi/B)^3} =
\frac{(B\sigma)^3<N>}{6\pi^2\eta},
\label{3.7}
\ee
$\eta=\frac{\pi}{6}\rho\sigma^3$ is fraction density.
Therefore, one may consider quantity B as the size of the first 
Brillouin zone of this block lattice.

The shift 
$$
\omk = \omk' + \Delta\delta_{\vk},
$$
$$
\rho_{\vk} = \rho'_{\vk} + \tilde \cM_1\delta_{\vk},
$$
  
where
$$
\Delta = -\frac{i}{2\pi}\frac{\bcM_3(0)}{\bcM_4(0)}\delta_{\vk},
$$
$$
\tilde \cM_1 = \bcM_1(0) -
\frac{\bcM_2(0)\bcM_3(0)}{\bcM_4(0)}+\frac{\bcM_3^3(0)}{3\bcM_4^2(0)},
$$
$\bigg (\bcM_n(0)=\frac{\cM_n(0)}{(\sqrt 2)^n}$, $n=1,...,4\bigg )$, 
transforms $\Xi$ into a form containing terms $\tilde \cM_1(0)$, $\tilde 
\cM_2(0)$ and $\tilde \cM_4(0)$ only (the primes on $\rhok$ and 
$\omega_{\vk}$ are omitted for clarity):
\bea
\Xi & = & \Xi_0 \Xi_G^{(1)} \int \exp \lbr \mu^*\rho_0 - \frac{\beta}{2}
  \sum_{\vk<B}\tilde V(k) \rhok\rhomk+\rd\non
& + & i2\pi \sum_{\vk<B} \omk\rhok - \frac{(2\pi)^2}{2}\tilde
  \cM_2(0)\sum_{\vk<B}
   \omk\omega_{-\vk} - \frac{(2\pi)^4}{4!<N_B>}|\tilde{\cM}_4(0)|\times\non
& \times & \ld
\sum_{\vk_1...\vk_4<B}\omega_{\vk_1}\omega_{\vk_2}\omega_{\vk_3}\omega_{\vk_4}
\delta_{\vk_1+...+\vk_4}
  \rbr (d\omega)^{N_B}(d\rho)^{N_B},
\label{3.8}  
\eea
where
\bea
\Xi_G^{(1)} & = & \Xi_G^c \exp \lbr \mu^*\tilde \cM_1 + \frac{\beta\tilde
  V^*(0)}{2} \tilde \cM_1^2 - \rd\non
& - &\ld\frac{\bcM_1(0)\bcM_3(0)}{\bcM_4(0)} - \frac{\bcM_2(0)\bcM_3^2(0)}{2\bcM_4^2(0)}-
   \frac{\bcM_3^4(0)}{8\bcM_4^3(0)}\rbr,\non
\mu^*& =& h-a_1,~~~a_1=\frac{\bcM_3(0)}{|\bcM_4(0)|}+\beta \tilde V^*(0)
\tilde
  \cM_1,~~~h = \beta \mu_1^+,\non
\tilde \cM_2(0) & = & \bcM_2(0) - \frac{\bcM_3^2(0)}{2\bcM_4(0)},\non
\tilde \cM_4(0) & = & <N_B> \bcM_4(0).
\label{3.9}
\eea
 Expression (\ref{3.8}) for $\Xi$ corresponds to the Ising model 
in the external field ($a_1-\beta\mu_1^+$) with the only difference: 
cumulants $\tilde\cM_2(0)$, $\tilde\cM_4(0)$ are functions of the fraction 
density $\eta$, temperature $T$ and parameters of the attractive interaction 
$\tilde\Phi_{\gamma\delta}(k)$.

After integration over $\omega_{\vk}$ we obtain the following form for 
the GPF: 
\bea
\Xi & = & \Xi_0\Xi_G^{(1)}\lb Z (\tilde \cM_2,\tilde \cM_4)\rb^{<N_B>} (\sqrt
  2)^{<N_B>-1}\times\non
 &\times & \int \exp [E_4(\rho)] (d\rho)^{N_B}.
\label{3.10}   
\eea
Here
\bea
E_4(\rho) & = & \mu^*\rho_0 - \frac{1}{2} \sum_{\vk<B} d_2(k) \rhok\rhomk
 - \frac{a_4}{4!<N_B>} \times\non
& \times & \sum_{\vk_1...\vk_4<B} \rho_{\vk_1}\ldots\rho_{\vk_4}
   \delta_{\vk_1+...+\vk_4} + \ldots~~~,
\label{3.11}
\eea

\bea
Z(\tilde \cM_2, \tilde \cM_4)  = \left( \frac{1}{2\pi}\right)^{1/2}
\left(\frac{3}{|\tilde \cM_4(0)|}\right)^{1/4}
  e^{x^2/4} U(0,x),\non
d_2(k) =  a_2+\beta\tilde V(k),~~~a_2=\sqrt{\frac{3}{|\tilde
\cM_4(0)|}}K(x),\non
a_4 =  \frac{3}{|\tilde\cM_4(0)|} L(x),
\label{3.111}   
\eea
where
\bea
K(x)  =  U(1,x)/U(0,x),\non
L(x)  =  3K^2(x) + 2xK(x) - 2,\non
x = \sqrt{\frac{3}{|\tilde \cM_4(0)|}} \tilde \cM_2(0).
\label{3.12}   
\eea
$U(a,x)$ is a parabolic cylinder function \cite{23}.
Expressions (\ref{3.10})-(\ref{3.12}) have the same forms as 
similar expressions for a one-component system obtained in \cite{17}. This 
coincidence is achieved due to the symmetry of the model under 
consideration. $E_4(\rho)$ is the Ginzburg-Landau-Wilson Hamiltonian for the 
SBFM in the vicinity of the vapour-liquid critical point. 

In order to integrate the GPF (\ref{3.10})-(\ref{3.12}) over 
$\rhok$ and determine the critical temperature we use the method developed 
in \cite{16,24} for the Ising model.
The essence of the method consists in subseqent integration over 
the layers of the CV space, beginning from $\rhok$ which correspond to 
short-wave fluctuations. Variations of the coefficients of $E_4(\rho)$ 
as the result of integration over $\rhok$ in $n$ subsequent layers of CV 
phase space are described by the recursion formulae derived in 
\cite{16}. For the case $T>T_c$ in the interval $[0,B]$ there 
exist three characteristic regions \cite{16}. The first region 
$B_{m_{\tau}}<k\leq B$ corresponds 
to the strongly correlated fluctuations $\rhok$, their density measure is 
non-Gaussian. The procedure based on the renormalization group 
symmetry is valid here. This is the region of the critical regime (CR). 
The second region $0<k\leq B_{m_{\tau}}$ is related to the fluctuation 
distributed according to the Gaussian density measure. This is the limiting 
Gaussian regime (LGR).

The third region consists of the point $k=0$. The variable $\rho_0$ is 
a macroscopic one and corresponds to the fluctuations of the particle 
density in the "external field" $\mu^*$.

We integrate (\ref{3.10}) according to the following scheme 
\cite{16}. The region $(0,B)$ is divided into the intervals $(B_1, 
B),...,(B_{i+1},B_i),...,$ where $B_n=B/S^n$ (S is a division parameter).
Each interval corresponds to a layer of subscripts $\vk$ in the Brillouin 
zone and each layer of subscripts $\vk$ - to a layer in the phase space 
$\rhok$. Integrating gradually over the layers we get a block lattice 
sequence with an appropriately growing block period and with the 
Hamiltonian corresponding to each block. Each Hamiltonian is characterized 
by the coefficients $d_2, a_4; d_2^{(1)}, a_4^{(1)}; d_2^{(2)}, a_4^{(2)}$, 
etc.. For the sequence of the block Hamiltonians  $\lbr d_2^{(n)}, 
a_4^{(n)} \rbr $  the renormalization group symmetry holds and the fixed 
point is of a saddle type. Because  the explicit expressions for the 
initial values of coefficients $d(k)$ and $a_4$ are given (see 
(\ref{3.11})-(\ref{3.12})), the solutions of the renormalization group 
type are functions of microscopic parameters, density and temperature.

The CR takes place for all the variables $\rhok$ at the critical 
point. Therefore, the critical temperature can be determined from the 
solution of recurrent equations. Using the result obtained in \cite{24} 
we derive the formula 
\bea
A(\beta_á\tilde V(0))^2+B(\beta_á\tilde V(0))+D=0,
\label{3.13}
\eea
wh¥re
\bea
A=1-f_0-R^{(0)}\sqrt{\varphi_0},\non
B=-a_2,\non
D=a_4R^{(0)}/\sqrt{\varphi_0}.\no
\eea
$f_0,\varphi_0$ are coordinates of the reduced fixed point, 
$R^{(0)}$ is a universal function of parameter S. The optimal value of
S is 3.4252 and the values of $f_0, \varphi_0, R^{(0)}$ corresponding to it 
are taken from \cite{24}. From the condition $\mu^*=0$ we obtain the 
second equation \cite{17}:
\be
\cM_3(0)=0,
\label{3.14}
\ee
which allows us to determine the critical density of the system. 

\section{Results and discussions}
In this section we present our results for the vapour-liquid critical point 
of symmetrical mixtures, using the method proposed above. These results 
are compared with those previously obtained by Monte-Carlo 
simulations \cite{13,14}.

The system under study is a symmetrical hard sphere square-well binary 
mixture. The interaction potential between the particles is given by:
$$
U_{\gs}(r)=\lbr
\begin{array}{cc}  \infty, & if~~~ r<\sigma\\
                  -\epsilon_{\gs}, & ~~~~~if~~ \sigma\leq r<\lambda\sigma\\

                                0, & if~~~ r\geq\lambda\sigma
\end{array}
\rd,~~~~~
$$
where $\sigma$ is a hard sphere diameter, $\lambda$ is a range of the 
potential, and $\epsilon_{\gs}$ is a well-depth of the interaction 
between the particles of types $\gamma$ and $\delta$. For a symmetrical 
mixture $\epsilon_{aa}=\epsilon_{bb}=\epsilon\neq\epsilon_{ab}$. In our 
formalism a completely analytical treatment for general $\lambda$ is 
possible. But in this paper we choose $\lambda=1.5$ and $\lambda=2.0$, 
because for these values of $\lambda$ the MC results are available. 

We split the potential $U_{\gs}(r)$ into short- and longe-range parts 
using the Weeks-Chandler-Andersen partition \cite{27}. As a result, 
we have: 
\bea
\Psi_{\gs}(r)=\lbr
\begin{array}{cc} \infty, & r\leq\sigma \\
                        0, & r>\sigma
   \end{array}
\rd,~~~~~
\label{3.1a}
\eea
\bea
\Phi_{\gs}(r)=\lbr
\begin{array}{cc} -\epsilon_{\gs}, & 0\leq r\leq\lambda\sigma\\
                        0, & r>\lambda\sigma
   \end{array}
\rd.~~~~~
\label{3.1b}
\eea
 
In our case the RS is a hard sphere system with the diameter 
$\sigma$ (see (\ref{3.1a})). The Fourier transform of function 
(\ref{3.1b}) has the form:
$$
\tilde \Phi_{\gs}(k)=\tilde \Phi_{\gs}(0) \frac{3}{(\lambda x)^3}[-\lambda x 
~cos(\lambda x) + sin(\lambda x)],\\
$$
where
$$
x=k\sigma,\\
$$
$$
\tilde \Phi_{\gs}(0)= -\epsilon_{\gs} \sigma^3 \frac{4\pi}{3} \lambda^3.
$$

Cumulants $\cM_{n}^{(i_n)}(0,...,0)$ are calculated according to the 
formulae given in Appendix B. For $S_2(0)$ the 
Percus-Yevick approximation is used:
\begin{displaymath}
S_2(0)= \frac{(1 - \eta)^4}{(1 + 2\eta)^2} .
\end{displaymath}

The solutions of equations (\ref{3.13})-(\ref{3.14}) are found 
numerically using a self-con\-sis\-tent procedure by means of which the 
dependences of the coefficients $a_2$ and $a_4$ (as well as cumulants 
$\cM_{n}(0)$) on $\beta_c$ are taken into consideration.

The vapour-liquid critical temperatures $T_c$ ($T_c=k_BT/\epsilon$) versus 
the microscopic parameter $r$ ($r=\epsilon_{ab}/\epsilon$ is a dissimilar 
interaction strength) are shown for $\lambda=1.5$ and 
$\lambda=2.0$ in figure 3.

\begin{figure}
\begin{center}
\centerline{
\epsfbox {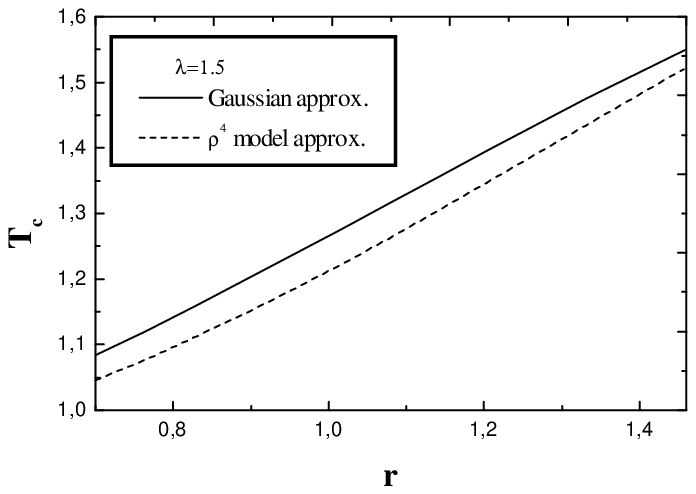}
\epsfbox {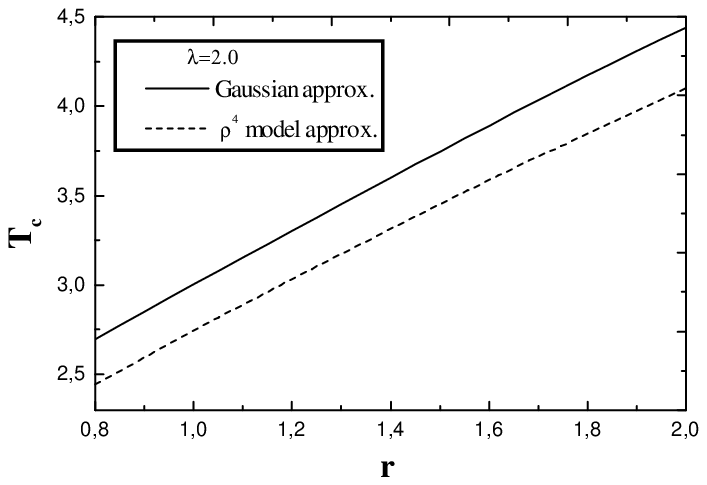}}
\end{center}
\caption{The 
vapour-liquid critical temperature as a function of the microscopic 
parameter $r$ at $\lambda=1.5$ (left) and $\lambda=2.0$ (right).}
\end{figure} 

In figure 4 we demonstrate the dependence of the 
critical density $\eta_c$ on $r$ for $\lambda$=1.5 and $\lambda$=2.0.
\begin{figure}[ht]
\begin{center}
\centerline{
\epsfbox {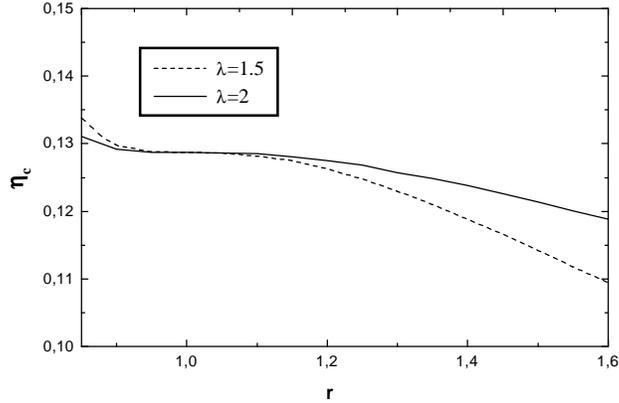}}
\end{center}
\caption{ The critical density as a function of the microscopic 
parameter r. }
\end{figure}
We compare our results with those obtained from MC simulations: for 
$\lambda=1.5$ and r=0.72 we have $T_c=1.055$, while the MC simulations give 
$T_c=1.06(1)$ \cite{14}, for $\lambda=2.0$ and r=1.0 (the case of 
r=1.0 corresponds to a one-component system) we obtain $T_c=2.753$ and 
$\eta_c=0.129$, while the simulations give $T_c=2.684(51)$ and 
$\eta_c=0.123(43)$ \cite{13}.

\section{Conclusions}
In this paper we propose a method 
for the study of the vapour-liquid critical point of a symmetrical binary 
mixture depending on its microscopic properties. Using this method we 
calculated the critical temperature and critical density of the symmetrical 
hard sphere square-well fluid versus the microscopic parameter $r$ which is 
the measure of the dissimilar interaction in the system. Our results agree 
with those obtained by MC simulations. We can improve our results in
the following ways: 1) taking into consideration the region of 
$\vk$ with $|\vk|>B$ (see figure 2); 2) using a higher approximation than 
the $\rho^4$ one.

\appendix
\section*{Appendix A}
\setcounter{section}{1}
A grand partition function of a two-component fluid system in  the
CV  representation with a RS can be written as in \cite{18}:

\begin{displaymath}
\Xi=\Xi_{0}\Xi_{1} ,
\end{displaymath}

where
\begin{displaymath}
\Xi_{0}=\sum_{N_{a}=0}^{\infty}\sum_{N_{b}=0}^{\infty}\prod_{\gamma=a}^{b}
\exp\left[\frac{\beta\mu_{0}^{\gamma}N_{\gamma}}{N_{\gamma}!}\right]
\int(d\Gamma)\exp\left[-\frac{\beta}{2}\sum_{\gamma,\delta=a,b}\sum_{i,j}
\psi_{\gamma\delta}(r_{ij})\right] 
\end{displaymath}
is a grand partition function of the RS; $\beta=\frac{1}{k_'T}$, $k_B$- is 
the Boltzman constant, $T$ is temperature;
 $(d\Gamma)=\prod_{a,b}d\Gamma_{N_{\gamma}}$,
$d\Gamma_{N_{\gamma}}=d\vec r_{1}^{\gamma}d\vec r_{2}^{\gamma}\ldots d\vec
r_{N_{\gamma}}^{\gamma}$
is a volume element of the configurational space of the $\gamma$-th species;
$\mu_{0}^{\gamma}$ is a chemical potential of the $\gamma$-th
species in the RS.

The part of the grand partition function which is  defined  in
the CV phase space has the form of a functional integral:

\be
\Xi_{1}=\int(d\rho)exp[\beta\sum_{\gamma}\mu_{1}^{\gamma}\rho_{0,\gamma}-
\frac{1}{2\beta}\sum_{\gamma,\delta=a,b}\sum_{\vk
}\alpha_{\gamma\delta}(k) \rho_{\vk,\gamma}\rho_{-\vk,
\delta}]J(\rho_{a},\rho_{b}) . 
\label{A.1}
\ee
Here,

1) $\mu_{1}^{\gamma}$ is a part of the chemical potential of the
$\gamma$-th species

\begin{displaymath}
\mu_{1}^{\gamma}= \mu_{\gamma}-\mu_{0}^{\gamma} +
\frac{1}{2\beta}\sum_{\vk}\alpha_{\gamma\gamma}(k),
\end{displaymath}

and is determined from the equation
\begin{displaymath}
\frac{\partial\ln\Xi_{1}}{\partial\beta\mu_{1}^{\gamma}} = \langle
 N_{\gamma}\rangle,
\end{displaymath} 
$\mu_{\gamma}$ is a full chemical potential of the $\gamma$-th
species; $\alpha_{\gamma\delta}(k)=\frac{\beta}{V} 
\tilde\phi_{\gamma\delta}(k)$ ; $<N_\gamma>$ is an average number of the 
$\gamma$-th species particles.

2)$\rho_{\vk,\gamma}=\rho_{\vk,\gamma}^{c}-i\rho_{\vk,\gamma}^{s} 
(\gamma=a,b)$ are collective variables of the $\gamma$-th species, the 
indices $c$ and $s$ denote the real part and the coefficient at the 
imaginary part of $\rho_{\vk,\gamma}$; $\rho_{\vk,\gamma}^{c}$ and 
$\rho_{\vk,\gamma}^{s}$  describe the value of the $\vk$-th fluctuation mode 
of the number of the $\gamma$-th species particles. Each  
$\rho_{\vk,\gamma}^{c}$ and $\rho_{\vk,\gamma}^{s}$ takes all the real 
values from $-\infty$ to $+\infty$. $(d\rho)$ is a volume element of the CV 
phase space:
\begin{displaymath}
(d\rho)=\prod_{\gamma}d\rho_{0,\gamma}{\prod_{\vk\not=0}}'
d\rho_{\vk,\gamma}^{c}d\rho_{\vk,\gamma}^{s}.
\end{displaymath}

The prime means that the product over $\vk$ is performed  in
the upper semi-space;

3) $J(\rho_{a},\rho_{b})$ is a transition Jacobian to the
CV averaged on the RS:
\begin{eqnarray}
J(\rho_{a},\rho_{b}) & = & \int(d\nu)\prod_{\gamma=a}^{b}\exp\left
[i2\pi\sum_{\vec k} \nu_{\vk,\gamma}\rho_{\vk,\gamma}\right]
\exp\left [\sum_{n\geq 1}
\frac{(-i2\pi)^{n}}{n!}\times\right.\nonumber\\ &  & \left.
\sum_{\gamma_{1}\ldots\gamma_{n}} \sum_{\vk_{1}\ldots\vk_{n}}
M_{\gamma_{1}\ldots\gamma_{n}}(\vk_{1},\ldots,\vk_{n}) 
\nu_{\vk_{1},\gamma_{1}}\ldots\nu_{\vk_{n},\gamma_{n}}\right] \no,
\label{A.3}
\end{eqnarray}
where  variables $\nu_{\vk,\gamma}$ are
conjugated to  CV $\rho_{\vk,\gamma}$.
$M_{\gamma_{1}\ldots\gamma_{n}}(\vk_{1}, \ldots,\vk_{n})$ is the
$n$-th cumulant connected  with
$S_{\gamma_{1}\ldots\gamma_{n}}(k_{1},\ldots,k_{n})$, the $n$-particle
partial structure factor of the RS, by means of the relation
\begin{displaymath}
M_{\gamma_{1}\ldots\gamma_{n}}(\vk_{1},\ldots,\vec
k_{n})= \sqrt[n]{N_{\gamma_{1}}\ldots
N_{\gamma_{n}}}S_{\gamma_{1}\ldots\gamma_{n}}
(k_{1},\ldots,k_{n})\delta_{\vk_{1}+\cdots+\vk_{n}},
\end{displaymath}
where $\delta_{\vk_{1}+\cdots+\vk_{n}}$ is a Kronecker symbol.

In general, the dependence of
$M_{\gamma_{1}\ldots\gamma_{n}}(\vk_{1}, \ldots,\vk_{n})$ on
wave vectors \\
$\vk_{1},\ldots,\vk_{n}$ is complicated. Hereafter we
shall replace $M_{\gamma_{1}\ldots\gamma_{n}}(\vk_{1}, \ldots,\vk_{n})$
by their values in the long-wave length limit
$M_{\gamma_{1}\ldots\gamma_{n}}(0,\ldots,0)$;

4)  $\tilde \phi_{\gamma\delta}(k)$ is a Fourier transform of the 
attractive potential $\phi_{\gamma\delta}(r)$. Function
$\tilde \phi_{\gamma\delta}(k)$ satisfies the following requirements:
$\tilde \phi_{\gamma\delta}(k)$ is negative for small values of $\vk$ and 
$lim_{\vk \to \infty}\tilde \phi_{\gamma\delta}(k)=0$. 

We pass in (\ref{A.1}) to CV $\rho_{\vk}$ and $c_{\vk}$
(according to $\omega_{\vk}$ and $\gamma_{\vk}$) by means of the
orthogonal linear transformation:
\bea
\rho_{\vk} =  \frac{\sqrt{2}}{2}(\rho_{\vk,a}+\rho_{\vk,b}),~~~
c_{\vk} =  \frac{\sqrt{2}}{2}(\rho_{\vk,a}-\rho_{\vk,b}),\non
\omega_{\vk} =  \frac{\sqrt{2}}{2}(\nu_{\vk,a}+\nu_{\vk,
b}),~~~
\nu_{\vk} =  \frac{\sqrt{2}}{2}(\nu_{\vk,a}-\nu_{\vk,b}).
\eea
Now $\rho_{\vk}$ and $c_{\vk}$ are connected with the total density 
fluctuation modes and the relative density (or concentration) fluctuation 
modes, respectively. 

As a result, for $\Xi_{1}$ we obtain formulae (\ref{2.2})-(\ref{2.6}).

\appendix
\section*{Appendix B}
\setcounter{section}{2}

Cumulants $\cM_n^{(i_n)}(0)$ with $n\leq 4$ are expressed in terms of 
the initial cumulants $\cM_{\gamma_1...\gamma_n}(0,...,0)$ 
$(\gamma_1,...,\gamma_n=a,b)$ as follows \cite{20}:
\bea
\cM_1^{(0)}(0)&=&\cM_a(0)+\cM_b(0)=<N>\non
\cM_1^{(1)}(0)&=&\cM_a(0)-\cM_b(0)=<N_a>-<N_b>\non
\cM_2^{(0)}(0)&=&\cM_{aa}(0)+\cM_{bb}(0)+2\cM_{ab}(0)\non
\cM_2^{(1)}(0)&=&\cM_{aa}(0)-\cM_{bb}(0)\non
\cM_2^{(2)}(0)&=&\cM_{aa}(0)+\cM_{bb}(0)-2\cM_{ab}(0)\non
\cM_3^{(0)}(0)&=&\cM_{aaa}(0)+\cM_{bbb}(0)+3[\cM_{aab}(0)+\cM_{abb}(0)]\non
\cM_3^{(1)}(0)&=&\cM_{aaa}(0)-\cM_{bbb}(0)+\cM_{aab}(0)-\cM_{abb}(0)\non
\cM_3^{(2)}(0)&=&\cM_{aaa}(0)+\cM_{bbb}(0)-\cM_{aab}(0)-\cM_{abb}(0)\non
\cM_3^{(3)}(0)&=&\cM_{aaa}(0)-\cM_{bbb}(0)-3[\cM_{aab}(0)-\cM_{abb}(0)]\non
\cM_4^{(0)}(0)&=&\cM_{aaaa}(0)+\cM_{bbbb}(0)+\non
   &+& 4[\cM_{aaab}(0)+\cM_{abbb}(0)]+6\cM_{aabb}(0)\non
\cM_4^{(1)}(0)&=&\cM_{aaaa}(0)-\cM_{bbbb}(0)+2[\cM_{aaab}(0)-\cM_{abbb}(0)]\non
\cM_4^{(2)}(0)&=&\cM_{aaaa}(0)+\cM_{bbbb}(0)-2\cM_{aabb}(0)\non
\cM_4^{(3)}(0)&=&\cM_{aaaa}(0)-\cM_{bbbb}(0)-2[\cM_{aaab}(0)-\cM_{abbb}(0)]\non
\cM_4^{(4)}(0)&=&\cM_{aaaa}(0)+\cM_{bbbb}(0)-\non
   &-& 4[\cM_{aaab}(0)+\cM_{abbb}(0)]+6\cM_{aabb}(0)
\eea
The same expressions hold at $\vk_i\neq 0$.

The $n$th cumulant $\cM_n^{(i_n)}(0)$ with $i_n=0$ is connected with the 
$n$th structure factor of the one-component system $S_n(0)$ \cite{20}: 
\begin{displaymath}
\cM_n^{(0)}(0)=<N>S_n(0).
\end{displaymath}

Structure factors $S_n(0) (n\geq 2)$ can be obtained from $S_2(0)$ 
by means of a chain of equations for correlation 
functions \cite{26}.
Cumulants with $i_n\neq 0$ can be exspressed 
in terms of $\cM_n^{(0)}(0)$ (see formulae (4.8) in \cite{20}).

\Figures
\Figure{Three phase regions of the symmetrical mixture depending 
on the microscopic parameters: (1) gas-gas and vapour-liquid phase 
transitions ($T_c^{g-g} > T_c^{v-l}$); (2) 
vapour-liquid and liquid-liquid phase transitions ($T_c^{v-l} > T_c^{l-l}$);
 (3) vapour-liquid phase transition only. $S_+$ is the structure factor of 
 the reference system. }
\Figure{The behaviour of the Fourier transform $\tilde 
V(k)/|\tilde V(0)|$ of the attractive part of the interaction potential 
V(r).}
\Figure{The 
vapour-liquid critical temperature as a function of the microscopic 
parameter $r$ at $\lambda=1.5$ (left) and $\lambda=2.0$ (right).}
\Figure{The critical density as a function of the microscopic 
parameter r.}

\end{document}